\documentclass[12pt]{article}

\usepackage{times}

\def\wt#1{\widetilde{#1}}
\def\ve{\varepsilon}

\def\ol#1{\overline{#1}}
\def\wh#1{\widehat{#1}}
\def\nn{\nonumber}
\def\eqn#1{(\ref{#1})}
 
\def\pa{\partial}

 \def\G{\Gamma}

\def\d{\partial} 
\def\e{\varepsilon}

 \def\L{\Lambda}
\def\m{\mu}
\def\n{\nu}

\def\r{\rho}
\def\s{\sigma}

\def\mn{{\mu\nu}}
\def\be{\begin{equation}}
\def\ee{\end{equation}}
\def\bea{\begin{eqnarray}}
\def\eea{\end{eqnarray}}

\begin{document} \thispagestyle{empty} \begin{flushright}
\framebox{\small BRX-TH~489 }\\
\end{flushright}

\vspace{.8cm} \setcounter{footnote}{0} \begin{center} {\Large{\bf
Stability of Massive Cosmological Gravitons}
    }\\[10mm]

{\sc S. Deser and A. Waldron
\\[6mm]}

{\em\small
Physics Department, Brandeis University, Waltham,
MA 02454,
USA\\ {\tt deser,wally@brandeis.edu}}\\[5mm]

\end{center}

\begin{center}{\sc Abstract}\end{center}

{\small
\begin{quote}
\noindent
We analyze the physics of massive spin~2 fields in (A)dS backgrounds
and exhibit that:
The theory is stable only for masses $m^2\geq2\L/3$, where 
the conserved energy associated with the background timelike 
Killing vector is positive, while  the instability for $m^2<2\L/3$
is traceable to the helicity 0 energy.
The stable, unitary, 
partially massless theory at $m^2=2\L/3$ describes 4 propagating
degrees of freedom, corresponding to helicities $(\pm2,\pm1)$ but 
contains no $0$ helicity excitation. 
\end{quote}}

\vspace{.5cm}

\noindent{{\small PACS numbers: \tt 03.65.Pm, 04.20.Fy, 04.40.-b, 04.62.+v}}

\section{Introduction}

Massive higher spin fields in cosmological, AdS ($\L < 0$) or dS
($\L > 0$) backgrounds have recently been shown to exhibit a novel
structure in the $(m^2,\L)$ plane~\cite{Deser:2001pe}, as compared to their
flat space counterparts where only the $m^2=0$ theory 
is distinguished.  The background space, with its
added parameter $\L$  affects the lower
helicity (in flat space language) modes in such a
way that they disappear entirely along
lines in the  $(m^2,\L)$ (half-)plane. Underlying the
appearance of these partially massless theories
are new gauge invariances. 
The lower helicities also flip from unitary to nonunitary as the
relevant lines are traversed. In particular, for massive spin 2
fields, it is known that the norm of the helicity zero mode
changes sign~\cite{Higuchi:1987py,Deser:2001pe} 
across the dS line $m^2=2\L/3$, 
along which a new local invariance appears~\cite{Deser:1983tm}.
The region $m^2<2\L/3$ is therefore unitarily forbidden.  

In this Letter we give a concrete proof of these results, {\it i.e.} 
that the $m^2=2\L/3$ partially massless spin~2 theory
describes 4 propagating degrees of freedom (PDoF)
corresponding to helicities $(\pm2,\pm1)$ (but {\it not} 0).
We then show that massive gravitons
are only stable in the unitarily allowed region
$m^2\geq2\L/3$. Stability in (A)dS is defined just as for
massless cosmological gravitons~\cite{Abbott:1982ff}, in terms of
positivity of the conserved energy associated with the timelike Killing vector
within the physically accessible
spacetime region, the intrinsic dS horizon.

In our
Hamiltonian (3+1) approach, the behavior of the various
helicity modes in the unitarily
allowed and forbidden $(m^2,\L)$ regions
and along the partially massless line is manifest.  
Since  massive spin~2 is described by
small oscillations of the cosmological Einstein theory
about its  vacuum, deformed by an explicit
mass term that breaks the linearized coordinate invariance of the former,
we utilize known aspects of the 
massless model~\cite{Abbott:1982ff}. 
The constraint structure and rich behavior in the
$(m^2,\L)$ plane of the massive model are, however,  very different.  
[Our stability analysis is carried out in a dS
background, but applies to AdS as well.]

In outline, we begin in Section~\ref{ADGCEA}
by writing down the 3+1 Hamiltonian representation of
the massive spin~2 theory in a dS background.
Away from the strictly massless $m^2=0$ (linearized cosmological
graviton) line, helicities $(\pm2,\pm1)$ are stable and
unitary since they are immune to the helicity 0 (scalar) constraint.
We derive their actions in Section~\ref{BEADFsB}.
The renegade 0 helicity, responsible for the non-unitary, unstable
region is analyzed in Section~\ref{CFAsDsGC}; we show both
that a novel constraint banishes
this excitation from the spectrum at $m^2=2\L/3$
and that the helicity 0 action goes from stable to unphysical as this
line is crossed. In Section~\ref{DGCFAD}, we 
map
the
stability regions of the models, and conclude with a brief discussion in
Section~\ref{dis}.

\section{The Action}

\label{ADGCEA}

We begin with the 3+1 form~\cite{Abbott:1982ff} 
of the cosmological Einstein action,
 \bea
  I_{E\L}&=&-\int d^4x\sqrt{-{}^{(4)}\!g}\ \Big[\ {}^{(4)}\!R-2\L\ \Big] \nn\\
         &=&\ \int d^4x\ \Big[\ \pi^{ij}\ \dot g_{ij}+N\  {\cal
         R}^0+N_i\ {\cal R}^i\ \Big]\ ,
 \label{smirnof}
 \eea
$${\cal R}^0=\sqrt{g}\, \,\Big[\ R-2\L\ \Big]
 +\pi^{ij}\pi^{lm}\,\Big[\ \frac{1}{2}\,g_{ij}g_{lm}-g_{il}g_{jm}
\ \Big]\Big/\sqrt{g} \ ,$$
$${\cal R}^i=2D_j\pi^{ij}\ , \qquad N\equiv \Big(-g^{00}\Big)^{-1/2}\, ,\quad
N_i\equiv g_{0i}\ .$$
Throughout, latin indices are spatial as are all derivatives and
index operations.  Our signature is
mostly plus, and the intrinsic
spatial Ricci tensor $R_{ij} \sim + \pa_k \G^k_{ij}$.
We expand~\eqn{smirnof} about its dS vacuum, using
the synchronous (if not fully covering) gauge
 \be
 ds^2=-dt^2+f^2(t)\ dx^i\, dx^i \, ,
 \ee
$$
f(t)\equiv e^{Mt}\ , \quad
M\equiv \sqrt{\L/3} \ .
$$
In this frame,  we will be almost able to remove all explicit
time dependence  due to $f$.  Denoting the full metric by $g_\mn$
and its above background value by $\ol{g}_\mn$, the deviations are
defined by
 \bea
  g_{ij}&=&\ol g_{ij}+h_{ij}=f^2\,\delta_{ij}+h_{ij} \, ,\nn \\
  \pi^{ij}&=&\ol\pi^{ij}+p^{ij}=-2M\,f\delta^{ij}+p^{ij} \, ,\\
N&=&1+n\, ,\quad g_{00}\equiv-1+h_{00}=-1+n/2+{\cal O}(n^2) \, .\nn
 \eea
Here $\ol{\pi}^{ij}$ is (essentially) the second fundamental form
in our gauge; $p^{ij}$ is of course the (independent) momentum conjugate
to $h_{ij}$; with respect to the background, $p^{ij}$ is a 
contravariant tensor
density, while $h_{ij}$ is a covariant tensor. The shift 
$N_i$ needs no expansion since its background value vanishes.

Before expanding the action~\eqn{smirnof}
to quadratic order in the deviations $(p^{ij},h_{ij}$, $N_i,n)$, 
we introduce the mass term. It maintains the background coordinate
invariance but breaks the linearized diffeomorphism symmetry of the
small oscillations, 
 \bea
  I_m&=&-\frac{m^2}{4}\,\int d^4x \sqrt{-^{(4)}\!\ol g}\;
        h_{\m\n}\ h_{\r\s}\ \Big[\ \ol g^{\m\r}\ol g^{\n\s}-\ol
        g^{\m\n}\ol g^{\r\s} \Big] \nn \\
     &=&-\frac{m^2}{4}\,\int d^4x \,f^{-1}\,
        \Big[\
        h_{ij}^2-h_{ii}^2-2f^2N_i^2-4f^2nh_{ii} 
        \ \Big]\ .
 \eea
In the last line (and from now on) we indicate the time dependence $f^{-1}$
explicitly and contract spatial indices with Kronecker deltas.
The massive spin~2 action in a dS background is, therefore,
\be
I=I_{E\L}^Q+I_m\, ,
\label{slat}
\ee
where
 \be
  I_{E\L}^Q=p^{ij}\, \dot h_{ij} +n\, {\cal R}^0_L+
  {\cal R}_Q^0+N_i {\cal R}^i_L\, .
 \ee
We denote an expression's  linear and quadratic parts
in the fluctuations $(p^{ij}$, $h_{ij}$, $N_i,n)$ by $L$
and $Q$, respectively. We also drop all integration signs and integrate freely
by parts. 

In absence of the
mass term (but with $\L \neq 0$), the four familiar constraints
${\cal R}^0_L = 0 = {\cal R}^i_L$, imposed by the (lapse and shift) 
Lagrange multipliers $n$ and
$N_i$, leave only the top, helicity $\pm$2,
linearized graviton excitations, 
consonant with the four gauge invariances of the
system~\cite{Abbott:1982ff}.
Addition of the mass term alters this counting: 
the $n^2$ term is still absent, but (for $m\neq0$) an $N^2_i$
term is present. Therefore only the $n$-constraint remains,
generically reducing the 6 canonical pairs $(p^{ij},h_{ij})$ 
to the five physical helicities $(\pm 2, \pm 1, 0)$ of massive spin~2.
However, as we shall demonstrate, along the line $m^2=2\L/3\equiv 2M^2$
a further constraint appears and excises the scalar helicity~0 mode. 

Assuming henceforth that $m^2\neq0$ (since the stability of
linearized cosmological gravitons is understood~\cite{Abbott:1982ff}), 
integrating out the shift function
$N_i$ yields the action
\be
I=p^{ij}\, \dot h_{ij} +n\, ({\cal R}^0_L+f^{-1}\,m^2h_{ii})
+{\cal R}_Q^0
-\frac{1}{2}\,f^{-1}\  
\Big(\frac{1}{m}\,{\cal R}^i_L\Big)^2-\frac{m^2}{4}\,f^{-1}\,
        \Big(h_{ij}^2-h_{ii}^2\ \Big)\ .
\label{FAsDsGsCF}
\ee

It is very convenient to minimize the explicit time
dependence (due to $f(t)$)
of the action, by making
the simple field redefinition
 \be 
  h_{ij}\longrightarrow f^{1/2} h_{ij}\, ,\qquad 
  p^{ij}\longrightarrow f^{-1/2} p^{ij}\, ,\qquad
  n\longrightarrow f^{-3/2} n \, .
 \label{splendid}
 \ee
The symplectic terms then become
 \be
  p^{ij} \ \dot h_{ij}\longrightarrow 
  f^{-1/2} p^{ij} \ \frac{d}{dt}\Big(f^{1/2} h^{ij}\Big)=
  p^{ij} \ \dot h_{ij}+\frac{M}{2}\,p^{ij} \ h_{ij}\, .
 \ee
It is easy to verify that the only remaining explicit time dependence of the
action is through the Laplacian 
\be
\nabla^2=f^{-2}\,\d_i^2\, .
\ee

Our analysis makes essential
use of the familiar flat 3-space orthogonal decomposition of 
symmetric 2-tensors,
 $$
  T_{ij}=T_{ij}^{Tt}+2\,\d_{(i}T^t_{j)}+\frac{1}{2}\,(\delta_{ij}-\wh\d_{ij})
         \,T^t+\wh \d_{ij}\,T^l\ ,   
 $$
 \be
  T_{ii}^{Tt}=0=\d_iT_{ij}^{Tt}=\d_iT_{i}^{t}\, ,\quad \wh
  \d_{ij}\equiv\d_i\d_j/\d_k^2 \ , 
 \label{EADGBE}
 \ee
which, of course, commutes with $\pa /\pa t$.  
The constraint ${\cal R}^0_L$, being a scalar linear in the fluctuations,
can only depend on the helicity 0, $(T^t,T^l)$ parts of
$(h_{ij},p^{ij})$.
Furthermore, since the action is of quadratic order,
there is no interaction between distinct helicities, schematically
 \be
 I = I_{\pm2}(T_{ij}^{Tt}) + I_{\pm 1}(T_i^t) + I_0(T^t,T^l)\ ,
 \label{cat}
 \ee
the hallmark of the orthogonal decomposition~\eqn{EADGBE}.
We now derive and examine each helicity term in turn.

 \section{Safe Helicities $(\pm2,\pm1)$}

\label{BEADFsB}

Helicities $(\pm2,\pm1)$ are the easiest part of the calculation
since they are unconstrained (for $m^2\neq0$).
Let us begin with the helicity
$\pm$2 part, where there is never a constraint.
We denote 
$(p^{Tt}_{ij},h^{Tt}_{ij})$ by $(p_{\pm2},q_{\pm2})$ respectively 
because, thanks to the transverse-traceless property, indices can only
contract in an obvious way\footnote{For example 
$p_{\pm2} q_{\pm2}\equiv\sum_{\e=\pm2}p_{\e}
q_{\e}=p_{ij}^{Tt}q_{ij}^{Tt}$.}.
By explicitly writing out the helicity $\pm2$ dependence of the
action~\eqn{FAsDsGsCF} [note that the
linearized Einstein tensor gives $G_L^{Tt}{}_{ij}=
-\frac{1}{2}\,\nabla^2 q^{Tt}_{ij}$] we find
 \be
  I_{\pm 2}  =  p_{\pm2}\, \dot q_{\pm2} - \Big[\
   \Big(p_{\pm2} +\frac{5M}{4}\,q_{\pm2}\Big)^2 \ + \
   \frac{1}{4}\; q_{\pm2}\, \Big(-\nabla^2+m^2
   -\frac{9M^2}{4}\Big)\,q_{\pm2} 
   \ \Big] \, .
 \ee
A field redefinition
 \be
 p_{\pm2}\longrightarrow \Big(p_{\pm2} -\frac{5M}{4}\,q_{\pm2}\Big)
                         \Big/\sqrt{2}\ ,\qquad
 q_{\pm2}\longrightarrow \sqrt{2}\; q_{\pm2}\ ,
 \label{putrile}
 \ee
yields the diagonal  action
 \be
 I_{\pm2} =  p_{\pm2}\, \dot q_{\pm2} 
           -\,\Big[\ \frac{1}{2}\, p_{\pm2}^2 \ + \ 
                   \frac{1}{2}\, q_{\pm2}\, \Big(-\nabla^2+m^2
                  -\frac{9M^2}{4}\Big)\,q_{\pm2}\ \Big] \ .
 \label{pm2}
 \ee
We will explain and meet again the effective mass $(m^2-9M^2/4)$ later,
and at present just reassure the reader that this action ensures 
stable, unitary propagation for {\it all} $m^2$.
Likewise, the string of field
redefinitions~\eqn{splendid} and~\eqn{putrile} is valid for any
$m^2$. Therefore the helicity $\pm2$ modes
propagate according to~\eqn{pm2} for all models in the $(m^2,\L)$
half-plane.

Next consider the transverse vector action, $I_{\pm 1}$.
The decompositions~(\ref{EADGBE},\ref{cat}) implies that the result takes
the form $\d_i T_j^t\  \d_{(i} T'_{j)}{}^t=-\frac{1}{2}\ 
T_i^t \,\d_j^2\,T'_i{}^t$,
which begs for the field redefinition
 \be
 h_i^t\longrightarrow q_{\pm1}\Big/\sqrt{-\d_j^2}\ ,\qquad
 p_i^t\longrightarrow p_{\pm1}\Big/\sqrt{-\d_j^2}\ ,
 \ee
(again we will suppress the sums over helicities $\pm1$).
Returning to the action~\eqn{FAsDsGsCF} and extracting its helicity
$\pm1$ dependence, after a somewhat lengthy computation\footnote{Since
$\sqrt{g}\,R$ is the usual Einstein action, its quadratic part is
$-\frac{1}{2}\,h_{ij} \, G^{ij}_L$, and hence does not contribute
in this sector, by the linearized Bianchi identity $\d_i G^{ij}_L=0$.}
we find
 \bea
  I_{\pm1}  &=&  2\,p_{\pm1}\, \dot q_{\pm1}\ 
  - \ \left[\ 2\,p_{\pm1}\,\Big(\,\frac{-\nabla^2+m^2}{m^2}\,\Big)\,p_{\pm1}
  \ -\  M\, p_{\pm1}\,\Big(\,\frac{-8\,\nabla^2+5m^2}{m^2}\,\Big)\,q_{\pm1}    
 \right.\nn\\&&\hspace{3.5cm}\left.
  \ + \ \frac{1}{2}\,q_{\pm1}\,\Big(\ m^2 + 4\ M^2\:
  \frac{-4\,\nabla^2+m^2}{m^2}\ \Big) \, q_{\pm 1}\,\right]\, .
 \eea
The field redefinition 
 \be
  q_{\pm1}\longrightarrow\frac{3M\,q_{\pm1}+2\,p_{\pm1}}{2m}\, ,\qquad
  p_{\pm1}\longrightarrow\frac{4M\,p_{\pm1}-(m^2-6M^2)\,q_{\pm1}}{2m}\, ,
 \label{GCFAsDG}
 \ee
yields the desired --stable and unitary-- action
 \be
 I_{\pm1} =  p_{\pm1}\, \dot q_{\pm1} 
           -\,\left[\ \frac{1}{2}\, p_{\pm1}^2 \ + \ 
                   \frac{1}{2}\, q_{\pm1}\, \Big(-\nabla^2+m^2
                  -\frac{9M^2}{4}\Big)\,q_{\pm1}\ \right] \ .
 \label{pm1}
 \ee
The helicity $\pm1$ action is identical to its 
$\pm2$ counterpart~\eqn{pm2} with one important difference:
The field redefinition~\eqn{GCFAsDG} is singular at $m^2=0$ 
(and complex for $m^2<0$). This reflects the gauge invariance 
at $m^2=0$ (and instability of the theory for $m^2<0$).  
The vector constraint, imposed by
the shift functions $N_i$, is reincarnated in the
$m^2=0$ theory and removes the above helicity $\pm1$ states.

 \section{Dangerous Helicity 0}

\label{CFAsDsGC}

For $m^2\neq0$,  helicities $(\pm2,\pm1)$ are
unaffected by constraints. The physical helicity 0 state
leads a more interesting
life as it can be (i) stable and unitary when $m^2>2\L/3\equiv2M^2$, 
(ii) absent when $m^2=2M^2$ or (iii) unstable and
nonunitary for $m^2<2M^2$. 
 
Before writing down an action for the helicity 0 excitations 
(analogously to the helicity $(\pm2,\pm1)$ ones in~\eqn{pm2}
and~\eqn{pm1}), we analyze the constraint 
imposed by
integrating out the lapse Lagrange multiplier $n$. 
Using $h_{ii}=h^t+h^l\ $ and writing out the linearization of ${\cal R}^0_L$
explicitly\footnote{We use the linearizations 
 $$
  \Big(\sqrt{g}\,R\Big)^L=-\nabla^2h^t\ ,\quad 
  \Big(\sqrt{g}\Big)^L=(h^t+h^l)/2\ , 
 $$
 $$
  \Big(\Big[\ \frac{1}{2}\,
  \pi^i{}_i{}^2-\pi_{ij}\pi^{ij}\Big]\Big/\!\sqrt{g}\Big)^L
  =-2M(p^t+p^l)+M^2(h^t+h^l)\ .$$} we obtain
 $$
  (-\nabla^2 + \n^2)\, h^t  + \n^2\,h^l - 2M \,(p^t+p^l)
  =0\; ,
 $$
 \be
  \n^2 \equiv (m^2 - 2M^2)\; .
 \label{curtailed}
 \ee
The sign of the
parameter $\nu^2$ controls the stability, unitarity and PDoF count 
of the model; negative values  will yield non-unitary,
unstable helicity 0 excitations.

Let us now examine the effect of the constraint~\eqn{curtailed} 
on the symplectic terms in the helicity 0 action
\be
I_0=p^l\ \dot h^l+\frac{1}{2}\,p^t\ \dot h^t -H_0(p^l,h^l;p^t,h^t)\, .
\ee
We choose (with no loss of generality in curved backgrounds) 
to eliminate the variable $p^t$
via~\eqn{curtailed}
\be
p^t=-p^l+\frac{1}{2M}\,\Big((-\nabla^2+\nu^2)\,h^t+\nu^2\, h^l\Big)\ ,
\ee
which leads to
\be
I_0=\Big(p^l-\frac{\nu^2}{4M}\,h^t\Big)\ 
\Big(\dot h^l-\frac{1}{2}\, \dot h^t\Big) 
-\frac{1}{4}\, h^t\  \nabla^2\  h^t
-H_0(p^l,h^l;h^t)\ .
\ee
Diagonalizing the kinetic terms by the field redefinition
\be
p^l\longrightarrow p_0+\frac{\nu^2}{4M}\,h^t\ , \qquad
h^l\longrightarrow q_0+\frac{1}{2}\,h^t\ ,
\ee
we are finally ready to display the full helicity 0 action
\bea
I_0&=&
\!p_0\,\dot q_0 
-  \left[\
    -\ \frac{3\ \nu^2\ m^2}{32M^2}\ (h^t)^2 
    \ -\ \frac{1}{2M}\, h^t \, 
    \Big(\ \nabla^2\  \Big[p_0-Mq_0\Big] 
    + \frac{\nu^2\ m^2}{4M}\, q_0\ \Big)\right.
\nn\\&&\qquad    \!\!\!\!\!\!  \!\left.
-\ \frac{2}{m^2}\, \Big[p_0-Mq_0\Big] \ \nabla^2 \ \Big[p_0-Mq_0\Big]
+\ \frac{3}{2}\, \Big[p_0-Mq_0\Big]\,\Big(
p_0-\frac{m^2}{3M}\, q_0\Big)\:
\right]\ .\nn \\ 
\label{full}
\eea
The denominators $M$ in this expression do not represent a genuine singularity,
but arise from choosing to solve the constraint~\eqn{curtailed} 
in terms of $p^t$.
In contrast, 
the denominators $m^2$ are due to integrating out
the shift $N_i$ and are a reminder (as we have seen already) of
the strictly massless $m^2=0$
gauge theory. The key point is to notice that the coefficient of
$(h^t)^2$ vanishes on the critical line $\nu^2=0$ (as well as at
$m^2=0$, concordant with the previous remark).
At criticality, the field $h^t$ appears only linearly and is a
Lagrange multiplier for a new constraint, whereas for $\nu^2\neq0$, we 
can integrate out $h^t$ by its algebraic field equation and there
are no further constraints. Let us deal with each of these cases in
turn.

\subsection{$\nu^2=0\ $: The Partially Massless Theory}

\label{star}

Consider the models with mass tuned to the cosmological constant via
$m^2=2\L/3$. As is clear from~\eqn{full}, the Lagrange multiplier $h^t$ 
imposes the new constraint
\be
p_0-M q_0=0\, .
\ee
Eliminating $q_0$ (say) and since $(p_0 \dot p_0)$ is a total derivative,
the 0 helicity action~\eqn{full} vanishes exactly,
\be
I_0=0\, .
\ee
It is known~\cite{Deser:1983tm} 
that the critical theory possesses a local 
scalar gauge invariance,
\be
\delta h_\mn = (D_{(\m}D_{\n)} + \frac{\L}{3}\,\ol  g_\mn )\, \xi(x)\, .
\ee
Thus, our result establishes that its effect is to remove the
lowest helicity excitation.
Therefore, the total action is
 \be
 I_{\nu^2=0} =  \sum_{\e=(\pm2,\pm1)}\, \left\{ p_{\e}\, \dot q_{\e} 
           -\,\Big[\ \frac{1}{2}\, p_{\e}^2 \ + \ 
                   \frac{1}{2}\, q_{\e}\, \Big(-\nabla^2
                  -\frac{M^2}{4}\ \Big)\,q_{\e}\ \Big]\right\} \ .
 \ee
[The effective mass $-M^2/4$ is the same one as in~\eqn{pm2} and~\eqn{pm1},
evaluated at $\nu^2=0$.]
These remaining helicity $(\pm2,\pm1)$ excitations are
both unitary~\cite{Higuchi:1987py,Deser:2001pe} and, as we will show,
stable.

\subsection{$\nu^2\neq0\ $: The Massive Theory}

We may now eliminate $h^t$ by its algebraic field equation
\be
 h^t 
    \ =\ -\frac{8M}{3\ \nu^2m^2}\,  
     \nabla^2\  \Big[p_0-Mq_0\Big] 
    - \frac{2}{3}\, q_0  \ ,
\ee
which yields the action
\bea
I_0&=&p_0 \dot q_0-
\left[\ 
\frac{1}{24M^2}\ \nu^2 m^2\, q_0^2 +
\frac{1}{6M}\,\Big[p_0-Mq_0\Big]\,\Big(
2\nabla^2-3m^2+9M^2\Big)\, q_0
\right.\nn\\&&\qquad\:\:\left.
+\  \frac{1}{6\n^2m^2}\,\Big[p_0-Mq_0\Big] 
\Big(4\nabla^4-12\n^2 \nabla^2+9\n^2m^2\Big) 
\Big[p_0-Mq_0\Big]\
\right]\; .\nn\\
\label{pre}
\eea
A penultimate field redefinition/canonical transformation
\bea
p_0&\longrightarrow& p_0+M
\left[q_0+\frac{2M}{\nu^2m^2}\,\Big(-2\nabla^2+3\nu^2-3M^2\Big)\ p_0\right]
\ ,\nn\\ \\
q_0&\longrightarrow& q_0+\frac{2M}{\nu^2m^2}\,
\Big(-2\nabla^2+3\nu^2-3M^2\Big)\ p_0\ ,
\eea
diagonalizes the action~\eqn{pre},
\be 
I_{0} =  p_{0}\, \dot q_{0} 
           -\,\left[\ \frac{1}{2}\,
\left[\!\frac{\nu^2\ m^2}{12M^2}\!\right]\, q_{0}^2 \ + \ 
                   \frac{1}{2}\,
\left[\!\frac{12M^2}{\nu^2\ m^2}\!\right]\, p_{0}\, \Big(-\nabla^2+m^2
                  -\frac{9M^2}{4}\Big)\,p_{0}\ \right] \ .
\label{pen}
\ee
Before we present the final, complete, action, some important comments on its
penultimate form~\eqn{pen} are needed:
\begin{itemize}
\item The sign of the parameter $\nu^2$ controls the positivity
of the Hamiltonian (and consequently the energy). Therefore we
find that the $(m^2,\L)$ plane is divided into a stable region $m^2\geq2\L/3$
and an unstable one $m^2<2\L/3$.
\item A final field redefinition, 
\be
p_0\longrightarrow -\ \frac{\nu m}{2\sqrt{3}\ M}\ q_0
\ , \qquad
q_0\longrightarrow \frac{2\sqrt{3}\ M}{\nu m}\ p_0
\label{last}
\ee
brings the helicity 0 action into the same form as its helicity $(\pm2,\pm1)$
counterparts~\eqn{pm2} and~\eqn{pm1}, but this is only legal in the 
stable massive region $m^2>2\L/3$.
\item The $\nu^2=0$ singularity signals the onset of a gauge invariance
where the constraint analysis of Section~\ref{star} must be applied.
\item The apparent singularity at $M=0$ is spurious and reflects our
(arbitrary) choice of solution to the constraint~\eqn{curtailed}. 
\end{itemize}
The final action for massive spin~2 in the 
region $m^2>2\L/3$ is
 \be
 I_{\nu^2>0} =  \sum_{\e=(\pm2,\pm1,0)}\, \left\{ p_{\e}\, \dot q_{\e} 
           -\,\Big[\ \frac{1}{2}\, p_{\e}^2 \ + \ 
                   \frac{1}{2}\, q_{\e}\, \Big(-\nabla^2 +m^2
                  -\frac{9M^2}{4}\ \Big)\,q_{\e}\ \Big]\right\} \ ,
 \ee
and describes stable, unitary, helicity $(\pm2,\pm1,0)$ excitations.

 \section{Stability: Positivity of the Energy}

\label{DGCFAD}

We are now ready to demonstrate the stability of the model
in the allowed region $m^2\geq 2\L/3$. The dS background
possesses a Killing vector
\be
\ol\xi^\m=(-1,Mx^i)\, , \qquad
\ol\xi^2=-1+\Big(fMx^i\Big)^2 \, ,
\ee
timelike
within the intrinsic horizon $(fMx^i)^2=1$. Therefore, in this region
of spacetime, it is possible to define a conserved energy whose
positivity guarantees the stability of the model.

Let us consider  helicity $\e$ 
(omitting 0 at criticality)
described by the conjugate pair
$(p_\e,q_\e)$, whose time evolution is generated by the Hamiltonian
\be
H_\e=\ \frac{1}{2}\, p_{\e}^2 \ + \ 
                   \frac{1}{2}\, q_{\e}\, \Big(-\nabla^2+m^2
                  -\frac{9M^2}{4}\ \Big)\,q_{\e}\ .
\label{pig}
\ee
Hence the field equations are\footnote{The resulting second order
field equation $-\ddot q_\e + 
\Big(\nabla^2 -m^2+\frac{9M^2}{4}\ \Big)\,q_{\e}=0$ agrees precisely
with the covariant one, $(D^2-m^2-2\L/3)\phi_{\m\n}=0$ (together   with
the usual onshell conditions $D.\phi_\n=0=\phi_\r{}^\r$) when written out
explicitly
in this frame, remembering the field redefinition~\eqn{splendid}.}
\be
\dot q_\e =p_\e\ , \qquad
\dot p_\e = \Big(\nabla^2
                  -m^2+\frac{9M^2}{4}\ \Big)\,q_{\e}\, .
\label{fe}
\ee
However, the Hamiltonian~\eqn{pig} is not conserved, thanks to 
the explicit time dependence $f^{-2}(t)$ in $\nabla^2$, which was to be
expected since it generates time evolution $\frac{d}{d t}$ rather than along
the Killing direction $\ol\xi^\m\d_\m$. 
Instead, the conserved energy is defined in terms 
of the stress energy tensor
\be
E_\e=T_\e^0{}_\m \ {\ol\xi}^\m = - T_\e{}^0{}_0
 + M\  x^i\  T_\e{}^0{}_i\, . 
\ee
The momentum density $T_\e{}^0{}_i$ will be defined below and
$-T_\e{}^0{}_0=H_\e$ is the Hamiltonian in~\eqn{pig}. 
For gravity, the momentum density $T^0_i$ is the  
quadratic part of the coefficient of $N_i$, and a similar result holds
here. Keeping track of our field redefinitions, we find
(modulo irrelevant superpotentials)
 \be
 T_\e{}^0{}_i = - p_\e\ \pa_i\  q_\e 
+ \frac{1}{2} \: \pa_i \ \Big(p_\e\, q_\e\Big)\, .
 \ee
It is not difficult (using~\eqn{fe} and spatial integrations by parts)
to verify that the energy function
\be
E_\e=H_\e-M\ x^i\  p_\e \pa_i  q_\e - \frac{3}{2}\ M 
\ p_\e q_\e \, ,
\ee
is indeed conserved, $\dot E=0$.

Finally we come to positivity. Here we need only a simple extension of
the method in~\cite{Abbott:1982ff}. Rewriting $E$ as
$$
E=\frac{1}{2}\,\Big(\wh x^i\,\wt p_\ve\Big)^2
+\frac{1}{2}\,\Big(f^{-1}\d_i\,q_\ve\Big)^2 -fM |x|\,\Big(\wh
x^i\,\wt p_\ve\Big)\,\Big(f^{-1}\,\d_i\,q_\ve\Big)
+\frac{1}{2}\,m^2\,q_\ve^2\, ,
$$
\be
\wt p_\e \equiv p_\e -\frac{3M}{2}\ q_\e\ , \qquad x^i\equiv |x|\,\wh
x^i\ ,
\label{triangle}
\ee
the first three terms
are positive by the triangle equality within the 
intrinsic dS horizon
\be
f\,M \,|x|<1\ ,
\ee
and the fourth, mass term is manifestly positive\footnote{
The Killing energy of a massive scalar  in dS also takes 
the form~\eqn{triangle} and is therefore stable for $m^2\geq0$.
[In this non-gauge example,
there is no analog to the spin~2 instabilities at negative values of
$m^2$ or $\nu^2$ whose vanishing is associated with 
gauge invariances.] Scalars in AdS actually enjoy a somewhat 
wider stability
range, extending to negative values of
$m^2$~\cite{Breitenlohner:1982bm} due to a shift in the spectrum
of the AdS 3-Laplacian. This broadening is unlikely for spin~2,
since its stability is  controlled entirely
by the above gauge coefficients.}
This concludes our
stability proof. 

The instability of the region $m^2<2\L/3$ is also manifest: 
Consider helicity $\e=0$.
Recall that
once $\nu^2<0$, we cannot make the rescalings with factors $\nu$ and
$\nu^{-1}$ in the final field redefinition~\eqn{last}.
This does not prevent us from constructing a conserved energy 
with a ``triangle'' form~\eqn{triangle}, but the {\it caveat}
is that  $p_0^2$ carries a factor $\nu^2$ and likewise $q_0^2$ a
factor $\nu^{-2}$. Therefore the
energy is negative and the theory is unstable in this region.
Clearly, helicity 0 is the sole felon responsible for this behavior.

\section{Discussion}

\label{dis}

Spin~2 excitations in (A)dS backgrounds have the following features
in the $(m^2,\L)$ half-plane:
(i) For generic $m^2$, there are 5 propagating helicities.
(ii) The $m^2=0$ strictly massless theory retains only helicity
$\pm$2 excitations thanks to the gauge invariance of small oscillations
of Einstein gravity about its (A)dS vacuum. These are both stable and
unitary. (iii) Between the vertical line $m^2=0$ and the dS line
$m^2 = 2 \L /3 $, all five helicities are
present, but the theory is both unstable and non-unitary
in the 0 helicity sector.
(iv) At the $m^2=2\L/3$ line,
a scalar gauge invariance allows both $\pm 2$, and $\pm 1$, but
removes the dangerous  helicity 0 excitation. This partially massless
model is unitary and stable. Furthermore, it is the unique spin~2
theory whose equation of motion
implies propagation along the null cone of the conformally flat dS
spacetime~\cite{Deser:1983tm}. 
(v) As $(m^2-2\L/3)$ turns positive
beyond this line, all five excitations behave and propagate
normally. This region includes all of AdS and
Minkowski space.

The splitting of the $(m^2,\L)$ half-plane into forbidden and allowed
regions separated by (partially) massless gauge lines occurs for
all spins $s>1$~\cite{Deser:2001pe} and it would be an amusing
exercise to carry out a Hamiltonian analysis for spin 3/2, to
exhibit the origin of the critical (AdS) line there; the required
formalism already exists \cite{Deser:1977ur}.

Another interesting question for higher spin theories is whether their
propagation is causal~\cite{Velo:1969bt}. Unitarity,
classical stability and causality are all directly related. 
As shown in~\cite{Deser:2000dz} (in a slightly different context), 
the failure of canonical commutators to 
support unitary representations also implies acausal propagation: The
spin~2 theory is acausal\footnote{The $s=2$ causality analysis in
Einstein spaces of~\cite{Buchbinder:2000fy} 
misinterprets the results of~\cite{Deser:1983tm} [as
well as of the incomplete Hamiltonian treatment of~\cite{Bengtsson:1995vn}] to
draw the erroneous conclusion that the partially massless theory is 
non-unitary, but points out correctly that propagation is causal. Indeed the
$m^2=2\L/3$ theory is the critical case where
propagation is null~\cite{Deser:1983tm}.} 
in the unstable, unitarily forbidden region.

\section*{Acknowledgments}

We thank A.\ Higuchi and I.\ Bengtsson for interesting
correspondence. This work was supported by the National Science
Foundation grant PHY99-73935.

\end{document}